\batchmode
\documentclass[12pt]{article}    
\usepackage[latin1]{inputenc}
\usepackage{hyperref}
\usepackage{graphics}
\usepackage{epsfig}
\usepackage{color}
\usepackage{slashbox}
\usepackage{amsmath}
\usepackage{rotating}
\usepackage[10pt]{moresize}
\usepackage[round,authoryear]{natbib}

\begin{document}
\newtheorem{definition}{Definition}
\newtheorem{theorem}{Theorem}
\newtheorem{example}{Example}
\newtheorem{corollary}{Corollary}
\newtheorem{lemma}{Lemma}
\newtheorem{proposition}{Proposition}
\newtheorem{remark}{Remark}
\newenvironment{proof}{{\bf Proof:\ \ }}{\qed}
\newcommand{\qed}{\rule{0.5em}{1.5ex}}
\newcommand{\bfg}[1]{\mbox{\boldmath $#1$\unboldmath}}

\newcommand{\fraca}[2]{\displaystyle\frac{#1}{#2}}

\def \B{{\rm I\kern -2.1pt B\hskip 1pt}}
\def \N{{\rm I\kern -2.1pt N\hskip 1pt}}
\def \R{{\rm I\kern -2.2pt  R\hskip 1pt}}
\def \E{{\rm I\kern -2.2pt  E\hskip 1pt}}

\thispagestyle{empty}

\begin{center}
\section*{Simulating Posterior Distributions for Zero--Inflated Automobile Insurance Data}
\end{center}
\begin{center}

{\sc \bf J.M. P\'erez S\'anchez$^a$ and E. G\'omez--D\'eniz$^b$\\

\noindent{\small\it$^a$  Department of Quantitative
Methods, University of Granada, Spain}\\
\small\it $^b$Department of Quantitative Methods in Economics and
T$i$DES Institute. University of Las Palmas de Gran Canaria,
Spain.}

\end{center}
\par\noindent
\begin{center}
{\bf Abstract}
\end{center}

\noindent Generalized linear models (GLMs) using a regression procedure to fit relationships between
predictor and target variables are widely used in automobile insurance data. Here, in the process of ratemaking and in order to compute the premiums to be charged to the policy--holders it is crucial to detect the relevant variables which affect to the value of the premium since in this case the insurer could eventually fix more precisely the premiums. We propose here a methodology with a different perspective. Instead of the exponential family we pay attention to the Power Series Distributions and develop a Bayesian methodology using sampling--based methods in order to detect relevant variables in automobile insurance data set. This model, as the GLMs, allows to incorporate the presence of an excessive number
of zero counts and overdispersion phenomena (variance larger than the mean).  Following this spirit, in this paper we present a novel and flexible zero--inflated Bayesian regression model. This model includes other familiar models such as the zero--inflated Poisson and zero--inflated geometric models, as special cases. A Bayesian estimation method is developed as an alternative to traditionally used maximum likelihood based methods
to analyze such data. For a real data collected from 2004 to 2005 in an Australian insurance company an example is provided by using
Markov Chain Monte Carlo method which is developed in WinBUGS package. The results show that the new Bayesian method performs
the previous models.

\vspace{0.2cm}

\noindent {\bf Keywords}: Automobile Insurance, Claim, Markov Chain Monte Carlo, Portfolio, Power Series Distribution,
Zero--Inflated.

\vspace{0.5cm}

\noindent {\bf Acknowledgements:} JMPS and EGD partially funded by grant ECO2013--47092 (Ministerio
de Econom\'ia y Competitividad).

\newpage

\begin{center}
\section*{Simulating Posterior Distributions for Zero--Inflated Automobile Insurance Data}
\end{center}
\begin{center}
{\bf Abstract}
\end{center}
\noindent Generalized linear models (GLMs) using a regression procedure to fit relationships between predictor and target variables are widely used in automobile insurance data. Here, in the process of ratemaking and in order to compute the premiums to be charged to the policy--holders it is crucial to detect the relevant variables which affect to the value of the premium since in this case the insurer could eventually fix more precisely the premiums. We propose here a methodology with a different perspective. Instead of the exponential family we pay attention to the Power Series Distributions and develop a Bayesian methodology using sampling--based methods in order to detect relevant variables in
automobile insurance data set. This model, as the GLMs, allows to incorporate the presence of an excessive number of zero counts and overdispersion phenomena (variance larger than the mean).  Following this spirit, in this paper we present a novel and flexible zero--inflated Bayesian
regression model. This model includes other familiar models such as the zero--inflated Poisson and zero--inflated geometric models,
as special cases. A Bayesian estimation method is developed as an alternative to traditionally used maximum likelihood based methods
to analyze such data. For a real data collected from 2004 to 2005 in an Australian insurance company an example is provided by using
Markov Chain Monte Carlo method which is developed in WinBUGS package. The results show that the new Bayesian method performs
the previous models.

\vspace{0.5cm}

\noindent {\bf Keywords}: Automobile Insurance, Claim, Markov Chain Monte Carlo, Portfolio, Power Series Distribution,
Zero--Inflated.

\section{Introduction and motivation}
There are a lot of applications involving discrete data for which
observed data show frequency of an observation zero significantly
higher than the one predicted by the assumed model. The problem of
high proportion of zeros has been an interest in data analysis and
modelling in many situations such as in the medical field,
engineering applications, manufacturing, economics, public health,
road safety, epidemiology and, in particular, actuarial data.
Models having more number of zeros significantly are known as
zero--inflated models. Poisson regression models provide a
standard framework for the analysis of count data. However, count
data are often overdispersed relative to the Poisson distribution.
One frequent factor of overdispersion is that the incidence of
zero counts is greater than expected for the Poisson distribution
and this is of interest because zero counts frequently have
special issue. For example, in counting claims from policyholders,
a policyholder may have no claims either because he/she is a good
driver, or simply because no risk factors have happened ``near''
his/her driving. This is the distinction between structural zeros,
which are (almost) inevitable, and sampling zeros, which occur by
chance. In the other hand, as it is pointed out by
\cite{denuitetal_2007}, overdispersion leads to underestimates of
standard errors and overestimates of Chi--square statistics and
this could derive in serious consequences. For example, some
explanatory variables may become not significant after
overdispersion has been accounted for.

In the last decades there has been considerable interest in models
for count data that allow for excess zeros, particularly in the
econometric literature. \cite{mullahy_1986} explores the
specification and testing of some modified count data models.
\cite{lambert_1992} provides a manufacturing defects application
of these models and discusses the case of zero--inflated Poisson
(ZIP) models. \cite{guptaetal_1996} provide a general analysis of
zero--inflated models. \cite{gurmu_1997} develops a semi--parametric
estimation method for hurdle (two--part) count regression models.
\cite{ridoutetal_1998} consider the problem of modelling count data
with excess zeros and review some possible models.
\cite{hall_2000} adapts Lambert's methodology to an upper bounded
count situation, thereby obtaining a zero--inflated binomial (ZIB)
model. \cite{ghoshetal_2006} introduce a flexible class of zero
inflated models which includes other familiar models such as the
Zero Inflated Poisson (ZIP) models, as special cases by using a
Bayesian estimation method. An overview of count data in
econometrics including zero inflated models is provided in
\cite{camerontrivedi_1998}. In insurance, \cite{yipandyau_2005} provide a
better fit to their insurance data by using zero--inflated count
models. \cite{boucheretal_2007} revise zero--inflated and hurdles
models with application to a Spanish insurance company and more
recently, \cite{mouatassimandezzahid_2012} analyze the
zero--inflated models with an application to private health
insurance data.

GLMs using a regression procedure to fit relationships between
predictor and target variables are widely used in automobile insurance data. Here, in order to compute the premiums to be charged to the policy--holders it is crucial to detect the relevant variables which affect to the value of the premium since in this case the insurer could eventually fix more precisely their premiums. We propose here a methodology with a different perspective. Instead of the exponential family we pay attention to the Power Series Distributions and develop a Bayesian methodology using sampling--based methods in order
to model an automobile insurance data set. This model, as the GLMs, led us to incorporate the presence of an excessive number of zero counts and overdispersion phenomena. Following this spirit, in this
paper we present a novel and flexible zero--inflated Bayesian
regression model. We compare several inflated and standard
models focused on applications in automobile
insurance. The Bayesian model proposed here includes other familiar models such
as the zero--inflated Poisson and zero--inflated geometric models,
as special cases. A Bayesian estimation method is developed as an
alternative to traditionally used maximum likelihood based methods
to analyze such data. For a real data collected from 2004 to 2005
in an Australian insurance company an example is provided by using
WinBUGS. The results show that the new Bayesian method performs
the previous models.

The structured of the paper is as follows. Section \ref{s2}
provides the zero--inflated power series distributions and the new
Bayesian model proposed here. Sections \ref{s3} looks at
automobile insurance application and \ref{s4} briefly concludes.

\section{Modelling zero--inflated data\label{s2}}
It is known that power series distributions form a useful subclass
of one--parameter discrete exponential families suitable for
modelling count data. From the original works of
\cite{kosambi_1949} and \cite{noack_1950} the power series
distribution has been very popular in the statistical literature
dealing with discrete distributions which belong to this simple
class. Two references concerning these features are \cite{patil_1962a} and
\cite{patil_1962b}. A revision of the power series distribution can be viewed in the
chapter two in \cite{johnsonetal2005}.

The probability function of the power series distribution becomes
\begin{eqnarray}
\Pr(X=x)=\frac{b(x) \theta^x}{f(\theta)},\quad x=0,1,\dots,\label{psd}
\end{eqnarray}
where $b(x)\theta^x\geq 0$, $b(x)$ is a function of $x$ or
constant, $f(\theta)=\sum_{x=0}^{\infty}b(x)\theta^x$ is
convergent and $\theta>0$ is refereed as the power parameter of
the distribution.  The family of discrete distributions defined in
(\ref{psd}) includes a broad class of known distributions, as the
Poisson, binomial, negative binomial, logarithmic series and the
Conway--Maxwell--Poisson distributions, among others. After
computing the probability generating function, given by
$G_X(z)=f(z\theta)/f(\theta)$, $|z|\leq 1$, it is simple to see
that the mean and variance of the power series distribution result
\begin{eqnarray}
E(X)=\mu &=& \frac{\theta f^{\prime}(\theta)}{f(\theta)},\label{mean}\\
var(X)=\sigma^2 &=& \frac{\theta^2 f^{\prime\prime}(\theta)}{f(\theta)}+
\mu(1-\mu).\nonumber
\end{eqnarray}

Thus the index of dispersion
\begin{eqnarray}
ID=\frac{\sigma^2}{\mu}=1+\frac{\theta f^{\prime\prime}(\theta)}{f^{\prime}(\theta)}-\mu\label{id}
\end{eqnarray}
accommodates for overdispersion when $\frac{\theta
f^{\prime\prime}(\theta)}{f^{\prime}(\theta)}-\mu>0$. For example,
when the Poisson distribution is considered we have that
$f(\theta)=\exp(\theta)$ and $ID=0$, i.e. we get equidispersion.
If $f(\theta)=(1-\theta)^r$, $r>0$, the distribution in
(\ref{psd}) reduces to the negative binomial distribution and from
(\ref{id}) we get that $ID=1+\theta/(1-\theta)>1$ and
overdispersion phenomena is obtained. Observe that for the
binomial and negative binomial cases, the corresponding additional
integer parameters, usually called $n>0$ and $r>0$, are considered
as nuisance parameters.

Starting with a distribution belonging to the Power series, a more
flexible model can be considered by adding a parameter which led
us to inflate the zero value of the empirical data when there
exists inflation of this.  Thus, zero--inflated power series
distribution contains two parameters. The first parameter $\omega$
indicates inflation of zeros and the other parameter $\theta$ is
that of power series distribution. A zero--inflated power series
distribution is a mixture of a power series distribution and a
degenerate distribution at zero, with a mixing probability
$\omega$ for the degenerate distribution. As
\cite{johnsonetal2005} point out, a very simple alternative for
modelling this setting is to add an arbitrary proportion of zeros,
decreasing the remaining frequencies in an appropriate manner. In
conclusion, zero--inflated models deal with the problem that the
data display a higher fraction of zeros (non--claims in our case)
and therefore appropriate for modelling counts that encounter
disproportionally large frequencies of zeros.

If we start with a discrete distribution $\Pr(Y=y)$, we can build
a zero--inflated distribution in a simple form (see
\cite{cohen1966}), by assuming
\begin{eqnarray}
\Pr(Y=y;\omega)=\left\{
\begin{array}{lr}
\displaystyle \omega+(1-\omega) \Pr(Y=0), & y= 0,\\
\displaystyle (1-\omega) \Pr(Y=y), & y\neq 0,
\end{array}
\right. \label{ZID}
\end{eqnarray}
\noindent where $\Pr(Y=y),\;x=0,1,\dots$, is the parent
distribution and
\begin{eqnarray}
-\frac{\Pr(Y=0)}{1-\Pr(Y=0)}<\omega<1.\label{support}
\end{eqnarray}

This last inequality allows the distribution to be well defined for certain negative values of $\omega$. The counterpart of this representation of the support of $\omega$ instead of the usual $0\leq \omega\leq 1$ is that the mixing interpretation is lost but, in practice the $\omega$ parameter can take negative values into the support given in (\ref{support}) and therefore (\ref{ZID}) is genuine. See for example \cite{bhattacharyaetal_2008}. Later we will see that this is the case for the data which it will considered here.

So, the probability mass function of the zero--inflated Power series distribution, $ZIPS (w, \theta)$,
results
\begin{eqnarray}
Pr(Y_i=y_i;\omega)= \left \{ \begin{array}{cc} \omega + (1-\omega)
 \displaystyle\frac{b(0)}{f(\theta)}, & y_i=0, \\
 (1-\omega)\displaystyle\frac{b(k) \theta^k}{f(\theta)}   & y_i=k\neq 0,
 \end{array}\right.
\end{eqnarray}
where $f(\theta)=\sum_{k=0}^{\infty} b(k) \theta^k$, $0 \le \omega
<1$ and $\theta>0$. The mean and the variance are
\begin{eqnarray}
E(y_i;\omega) &=& (1-\omega)\mu,\\
var(y_i;\omega) &=& (1-\omega)(\sigma^2+\omega \mu^2),
\end{eqnarray}
where $\mu$ denotes the mean of the power series distribution given in
(\ref{mean}).

Now, zero--inflated forms assuming different count distributions
belonging to the power series distribution, can be defined easily.
\cite{guptaetal_1996} and \cite{ghoshetal_2006}, for example,
investigated the zero--inflated form of the generalized Poisson
distribution.

Maximum likelihood estimators of $\omega$ and $\theta$ can be
obtained by maximizing  $\log\ell(\omega,\theta;y_i)$,
$y=1,\dots,n$, with respect to $\omega$ and $\theta$, where
\begin{eqnarray}
\ell(\omega,\theta;y_i)=\prod_{i=1}^{n}\left[\omega + (1-\omega)
 \displaystyle\frac{b(0)}{f(\theta)}\right]^{n_0}
 \left[(1-\omega)\displaystyle\frac{b(y_i) \theta^{y_i}}{f(\theta)}
 \right]^{\alpha_i}.\label{loglike}
\end{eqnarray}

Here, $n$ is the sample size and $n_0$ is the number of zeros
counts in the sample. Observe that by using binomial expansion the
likelihood function in (\ref{loglike}) can be written as
\begin{eqnarray}
\ell(\omega,\theta;y_i)\propto
\theta^{n\bar x}\sum_{j=0}^{n_0}{n_0\choose j}\omega^j(1-\omega)^{n-j}
\left[\frac{b(0)}{
f(\theta)}\right]^{n-j}.\label{loglike1}
\end{eqnarray}

After obtaining the normal equations we have to solve the equation
\begin{eqnarray*}
\bar x-\frac{\theta f^{\prime}(\theta)}{f(\theta)-b(0)}=0
\end{eqnarray*}
to get the maximum likelihood estimate of $\theta$ and where $\bar
x$ is the sample mean. Once obtained $\theta$ the parameter
$\omega$ is obtained from
\begin{eqnarray*}
\omega=\frac{(n-n_0)f(\theta)}{n[f(\theta)-b(0)]}.
\end{eqnarray*}

Therefore, the maximum likelihood estimation of the parameters
under the power series distribution is simple and, in a similar
manner, the regression coefficients when covariates are implemented
into the model can also be obtained in a simple way.

\subsection{Including covariates}
In practice, the practitioner usually uses data set with commonly
available exogenous covariates in order to explain the variable
$Y_i$, known in this case as the endogenous variate. That is,
suppose that for the $i$th observation, covariates $x$ and $z$ are
available. In order to adapt the model to this framework we need
to relate these covariates with endogenous variable via the
parameters $\theta$ and $\omega$. This can be made through the
following links,
\begin{eqnarray*}
\theta_i &=& \exp(x^\top \beta),\\
\log\left(\frac{\omega_i}{1-\omega_i}\right) &=& z^{\top}\gamma
\end{eqnarray*}
with $\beta^{\top}=(\beta_1,\dots,\beta_k)$ and
$\gamma^{\top}=(\gamma_1,\dots,\gamma_k)$ vectors of unknown
regression parameters associated with covariates. Of course that
in practice is common to suppose that the design matrix $X$ and
$Z$ are the same.

A nice reformulation of the zero--inflated model above was
proposed recently by \cite{ghoshetal_2006} which considered that
the zero--inflated model can be represented as $Y= V(1-B)$, where
$B$ is a Bernoulli, Bernoulli($p$), random variable and $V$
independently to $B$ has a discrete distribution on the power
series, PS($\theta$). Under this representation, the mean, $E(Y)$,
and variance, $var(Y)$, can be rewritten as
\begin{eqnarray}
E(y) &=& (1-\omega) E(V), \\
var(y) &=& \displaystyle\frac{\omega}{1-\omega} E(y)^2 + \delta E(y),
\end{eqnarray}
where $\delta=var(V)/E(V)$ denotes the coefficient of dispersion
of the latent random variable $V$. If the latent variable, $V$
does not have an underdispersed distribution (i.e., $\delta \ge
1$), then the distribution of $Y$ is overdispersed. On the other
hand, if $V$ has a underdispersed distribution (i.e., $\delta<1$),
then $Y$ has a underdispersed distribution if and only if $E(V)<
(1-\delta)/w$. In their paper, they suppose that $V$ follows the
power series distribution. In advance, this model will be called
BZIPS (Bayesian Zero--Inflated Power series model).

In this paper two discrete distribution belonging to the Power
series distributions will be considered. They are the Poisson
distribution with parameter $\theta>0$ and the geometric
distribution with parameter $1/(1+\theta),\;\theta>0$.

\subsection{The Bayesian model}

In this section a Bayesian methodology is carried out which allows
for estimating of the model above in a simple way facilitating the
process to incorporate covariates and providing exact posterior
inference up to a Monte Carlo error. This model can easily
accommodate multiple continuous and categorical predictors.

From a Bayesian point of view, prior distributions for $\omega$
and $\theta$ will be required. In this sense and looking to the
log--likelihood in (\ref{loglike1}) it is adequate to assuming a
Beta prior distribution for $\omega$ and the natural conjugate
prior (\cite{ghoshetal_2006}) for the power series distribution in
the following way,
\begin{eqnarray*}
\omega & \sim & Be(b_1,b_2),\\
\theta & \sim & \pi(\theta)=\frac{\theta^{a_1}}{[f(\theta)]^{a_2}}.
\end{eqnarray*}

Both assumptions establish a congruent model, present important
computational advantages and, in addition, have a tradition in
Bayesian statistical literature. However, in this article, the covariates that
affect $\omega$ and $\mu$ are fixed. So, we specify independent prior distributions for
the parameters of the regression models, i.e. $\beta$ and $\gamma$, in the next way,
\begin{eqnarray*}
\beta & \sim & U(a_{\beta},b_{\beta}),\\
\gamma & \sim & U(a_{\gamma},b_{\gamma}),
\end{eqnarray*}
\noindent where the constants $a_{\beta}$, $b_{\beta}$,
$a_{\gamma}$ and $b_{\gamma}$ are assumed known. In particular,
$a_{\beta}=a_{\gamma}=-10^5$ and  $b_{\beta}=b_{\gamma}=10^5$
which expresses lack of knowledge about the regression parameters.
These noninformative uniform distributions are appropriate if no
prior knowledge is available about the likely range of values of
the parameters (\cite{lempers_1971};
\cite{mitchellandbeauchamp_1988}; among others).

Given that the prior distributions for parameters have been
assessed, the next procedure is combine the likelihood function in
(\ref{loglike1}) with priors to make a Bayesian inference. Since
no closed forms are available for marginal posterior
distributions, numerical approaches have to be used for generating
them. The numerical approaches used are based on simulating from
the posterior distributions which are proper since we consider
proper priors although with somewhat large variances. The
simulation approach used is Markov Chain Monte Carlo, which can be
done using the WinBUGS software. Markov Chain Monte Carlo (MCMC)
is in this setting a powerful tool which allows to get the
estimates of the parameters involved. As it is well--known, MCMC is
a method of posterior simulation and led us to compute the
posterior density function for arbitrary points in the parameter
space. With MCMC, it is possible to generate samples from an
arbitrary posterior density and to use these samples to
approximate expectations of quantities of interest such as the
mean or second order moment. Several other aspects of the Markov
chain method also contributed to its success. The methodology is
very simple and consist of generating simulated samples from that
posterior density, even though the density corresponds to unknown
distributions. In this context, Gibbs sampling is a natural
estimation method. A reasonable choice for starting values of
$\beta$ and $\gamma$ for the MCMC simulation can be obtained by
standard Poisson and Negative Binomial regression models using any
statistical software package such as STATA. In this work, all
simulations were done using WinBUGS
(\cite{spiegelhalteretal1999}). We run three parallel chains and a
single long chain for diagnostic assessment (checked using CODA
software). A total of 100,000 iterations were carried out (after a
100.000 burn--in period). A complete Gibbs sampling algorithm is
outlined in \cite{ghoshetal_2006}.

\section{Experiments with insurance data\label{s3}}
In this section an application to the different models considered
in the previous sections is developed in order to see how the
proposed Bayesian method works. The data set considered was taken
from the web page of MacQuarie University in Sydney (Australia) in
which the different data taken by \cite{jongandandheller_2008} are
available. This page contains numerous data to be used in
actuarial setting.

\subsection{The data}
In particular, the data studied contain information from policyholders of several
Australian insurance companies in 2004 and 2005, describing
certain characteristics related to the vehicles and the
policyholders. It contains 67856 policies, of which 63232
(93.18$\%$) has no claims. Table~\ref{descriptive} shows a
descriptive of the dependent and independent variables.

\begin{table}[htbp]
\caption{Descriptive summary of variables.\label{descriptive}}
\begin{center}
\begin{tabular}{lcccr}
\hline
Variables & Mean & Variance & Minimum & Maximum \\
\hline
Number of claims & 0.07275 & 0.07739 & 0 & 4 \\
Vehicle value & 1.77702  & 1.45258 & 0 & 34.56 \\
Gender  & 0.43110 & 0.24525 & 0 & 1 \\
Young age & 0.27436 & 0.19908 & 0 & 1 \\
Medium age & 0.57492 & 0.24439 & 0 & 1 \\
Old age & 0.46081 & 0.24846 & 0 & 1 \\
Vehicle age  & 0.57492 & 0.24439 & 0 & 1 \\
\hline
\end{tabular}
\end{center}
\end{table}

Table \ref{index} shows some measures which led us to consider
departures from the Poisson distribution. These measures are the
proportion of zeros, $p_0$, the cumulant $\kappa_3$, the
zero--inflation index, $z_i=1+\log(p_0)/\mu$, and the third
central moment inflation index, $\kappa=\kappa_3/\mu-1$, (see
\cite{puigvalero_2006} for details). In this Table we can see the
sample values of them and its corresponding estimates values
obtained by using the Poisson distribution, the geometric
distribution, the zero--inflated Poisson and the zero--inflated
geometric distributions after estimating the corresponding
parameters by maximum likelihood method. We can see that the
geometric distribution (in its versions inflated and non--inflated
at zero) outperforms the Poisson one.
\begin{table}[htbp]
\caption{Some measures of inflation\label{index}}
\begin{center}
\begin{tabular}{lccccc}\hline
 & Sample & Poisson & Geometric & ZIP & ZIG \\ \hline
 $p_0$ & 0.93180 & 0.92983 & 0.93218 & 0.93180 & 0.93186 \\
 $\kappa_3$ & 0.08757 & 0.07276 & 0.08941 & 0.08573 & 0.08705\\
 $z_i$ & 0.03000 & 0.00000 & 0.03470 & 0.03000 & 0.03000\\
 $\kappa$ & 0.20367 & 0.00000 & 0.22886 & 0.17832 & 0.19640\\

\hline
\end{tabular}
\end{center}
\end{table}

For each policy, the initial information for the period considered
and the existence or otherwise of at least a claim were reported
within this yearly period. In total, four explanatory variables
are considered, together with a dependent variable representing
the number of claims. Vehicle value represents its value in 10.000
Australian dollars. Vehicle age is equal to 1 if the vehicle is
relatively young (7 years or less). Gender is equal to 1 if the
policyholder is a man. This variable is included in the model for
didactic purposes but, as expected, is not relevant in all the
models considered. Finally, a categorical variable was considered
to represent the age of the policyholder by dividing this feature
in three dummies: young, medium and old ages. In this sense, we
try to identify if there is/are age sets with more propensity to
make claims. Several authors have used previously age variable in
a dichotomous way. Some of them are \cite{boucheretal_2007},
\cite{bermudez_2009} and \cite{perezetal_2014}, among others.

\subsection{Results, diagnostic and comparisons}
Table~\ref{zips} shows results under the Bayesian model proposed
by \cite{ghoshetal_2006} (BZIPS). As we can see, vehicle value and older
policyholders (in relation to medium age) are relevant for the
chance of being in zero--state. The positive value of the first
coefficient indicates that this chance increases with the value of
the vehicle (with relevance at 1$\%$) and the negative scope of
old age indicates that the change of being in zero--states
decreases for older policyholders in relation to the medium
age policyholders, at 5$\%$ of relevance.

In the other hand, a negative intercept of --1.883 (with relevance at 1$\%$) indicates that the average number of claims is less for the medium age policyholders. Furthermore, the more vehicle value, less average number of claims is expected, again at 1$\%$ of relevance. These results are consistent with the previous zero--state coefficients.

\begin{table}[htbp]
\caption{Zero inflated Power Series fitted model (BZIPS).\label{zips}}
\begin{center}
\begin{tabular}{lll}
\hline
 & ZIPS ($\alpha_i$) & ZIPS ($\beta_i$) \\
\hline
Intercept & --0.482 & --1.883 *** \\
          & (0.295) & (0.122) \\
Vehicle value & 0.543 *** & --0.102 *** \\
              & (0.096)  & (0.026) \\
Gender  & 0.005 & --0.028 \\
              & (0.181) & (0.077) \\
Young age  & --0.0008 & 0.095 \\
              & (0.237)  & (0.093) \\
Old age  & --0.497 ** & 0.018\\
              & (0.212) & (0.101)  \\
Vehicle age  & 0.070 & -0.035\\
              & (0.236) & (0.097)  \\
\hline

\end{tabular}
\end{center}
\end{table}

Table~\ref{zipsgeo} shows results under the Bayesian model proposed in this paper (BZIGPS).
Now, the fact of being in the medium age class increases the chance of zero--state (the $\alpha$--intercept is relevant at $10\%$). The average number of claims increases if the vehicle value is high (at 1$\%$ of relevance) and decreases in the medium and old age classes (more for older policyholders). For the youngest people, the average number of claims is expected to be greater than for the other two groups of policyholders (with $1\%$ of relevance). Finally, the more vehicle age decreases the average number of claims at 10$\%$ of relevance.

\begin{table}[htbp]
\caption{Zero inflated Geometric--Power Series fitted model (BZIGPS).\label{zipsgeo}}
\begin{center}
\begin{tabular}{lll}
\hline
 & BZIGPS ($\alpha_i$) & BZIGPS ($\beta_i$) \\
\hline
Intercept & 5.142 * & --2.637 *** \\
          & (3.199) & (0.051) \\
Vehicle value & 3.097 & 0.046 *** \\
              & (3.328)  & (0.022) \\
Gender  & 2.055 & --0.020 \\
              & (3.962) & (0.032) \\
Young age  & 0.311 & 0.096 ***\\
              & (4.628)  & (0.040) \\
Old age  & 2.421 & --0.211 *** \\
              & (4.705) & (0.043)  \\
Vehicle age  & 2.672 & --0.054\\
              & (4.441) & (0.037) *  \\
\hline

\end{tabular}
\end{center}
\end{table}

Although there are a variety of methodologies to compare
several models for a given data set, the deviance information
criterion DIC, which is a generalization of the AIC (Akaike
information criterion) and BIC (Bayesian information criterion),
is useful in Bayesian model selection problems where the posterior
distributions of the models have been obtained by MCMC simulation.
Recall that it is only valid when the posterior distribution is
approximately multivariate normal, which is the case considered
here. In all the criterions, the model that better fits is a data set is the model with the smallest
value.

Finally, it is interesting to compare these values with the
frequentist point estimates and standard errors. Table \ref{stan}
shows the results under two standard Poisson and negative binomial distributions and their zero--inflated versions.
We can observe that they detect the same relevant factors than
BZIGPS model but, as we can see in Table \ref{statist}, these models fits worse than the BZIPS
and BZIGPS models. The deviance of BZIPS is the
smallest, but the confidence interval with a 5$\%$ significance is
$[ 9336, 9889 ]$, which includes the deviance of the BZIGPS model,
so there is no significative difference in terms of fitting between
these two models. By comparing Tables \ref{zipsgeo} and \ref{stan}, we can observe that
the estimated coefficients differ considerably, although the signs and the relevant factors remain the same.
So, we can say that results from BZIGPS model are more consistent (than BZIPS)
with respect to the results from classical models in Table \ref{stan} providing much
better fitting.

\begin{table}[htbp]

  \resizebox{0.90\textwidth}{!}{\begin{minipage}{\textwidth}
        \caption{Poisson and negative binomial inflated and non--inflated
models\label{stan}}
\begin{center}
\begin{tabular}{lcccc}
\hline
 & Poisson & Negative Binomial & ZIP & ZIBN  \\
\hline
Intercept & -2.6282 ***  & -2.6305 ***  & -2.0540 ***   &  -2.6306 *** \\
          & (0.0407)   & (0.0412) & (0.0696) & (0.0412) \\
Vehicle value & 0.0381 *** & 0.0392 *** & 0.0392 *** & 0.0393 *** \\
              & (0.0114)   & (0.0117) & (0.0118) & (0.0117) \\
Gender  & --0.0159 & --0.0161 & --0.0164 & --0.0162 \\
              & (0.0300)   & (0.0300) & (0.0301) & (0.0301) \\
Young age & 0.0958 *** & 0.0955 *** & 0.0949 *** & 0.0956 *** \\
              & (0.0337)   & (0.0337) & (0.0337) & (0.0337) \\
Old age & --0.2081 *** & --0.2083 *** & --0.2082 *** & --0.2083 *** \\
              & (0.0385)   & (0.0385) & (0.0385) & (0.0385) \\
Vehicle age  & --0.0617 * & --0.0606 *  & --0.0604 * & --0.0606 * \\
              & (0.0334)   & (0.0335) & (0.0335) & (0.0335) \\
              \hline
Inflation constant &       &          & --0.2489 * & --14.1262 *** \\
                   &       &          & (0.1283) & (0.7096) \\

\hline
\end{tabular}
\end{center}
\end{minipage}}
\end{table}

\begin{table}[htbp]
\caption{Summaries of the fitting values for the models
considered: DIC, AIC and BIC\label{statist}}
\begin{center}
\begin{tabular}{lrrr}
\hline
 & DIC & AIC & BIC  \\
  \hline

 Poisson & 36118.353 & 36130.353 & 36185.105 \\
 Negative Binomial & 36019.507 & 36033.507 & 36097.383 \\
 ZIP & 36026.881 & 36040.881  & 36104.757\\
 ZINB & 36019.507 & 36035.507 & 36108.508 \\
 BZIGPS & 9871.000 & 9893.000  & 10004.501\\
 BZIPS & 9611.000 & 9623.000  & 9744.502\\

\hline
\end{tabular}
\end{center}
\end{table}

\section{Final remmarks\label{s4}}
Insurance literature have paid a lot of attention on the considerable potential of generalized
linear models as a comprehensive modelling tool for the study of the
claims process in the presence of covariates in automobile insurance. Nevertheless, less attention have been paid to the well--known family of Power Series Distribution.

In this paper we develope a Bayesian methodology using sampling--based methods in order to model an automobile insurance data set by using discrete distributions belonging to the Power Series Distributions. As a consequence, we get a new and flexible model when
overdispersion and inflation of zeros is presented in the data
set. This is a topic which frequently appears in data related with automobile insurance.

For a real data collected from 2004 to 2005
in an Australian insurance company an example is provided by using
Markov Chain Monte Carlo method which is developed in WinBUGS
package. Comparisons with
standard and Bayesian ZIP models in terms of parameter estimations and information
criterions such as DIC, AIC and BIC are carried out. Bayesian ZIP models suggest a much improved
fit over the classic ZIP models. Therefore, the results obtained here show that the new Bayesian method performs
the previous GLMs models.

\bibliographystyle{apalike}

\end{document}